\documentclass[12pt]{article}

\input epsf
\usepackage{latexsym}
\def\beq{\begin{eqnarray}}
\def\eeq{\end{eqnarray}}

\def\t{\widetilde}

\begin {document}

\begin{titlepage}
\begin{flushright}
{ YITP-SB-07-01 }
\end{flushright}
\vskip 0.9cm

\centerline{\Large \bf On Semiclassical Limits of String States}
\vspace{0.5cm}

\vskip 0.7cm
\centerline{\large Jose J. Blanco-Pillado$^{a,}$\footnote{E-mail:
jose@cosmos.phy.tufts.edu}, 
Alberto Iglesias$^{b,}$\footnote{E-mail: iglesias@physics.ucdavis.edu}
and Warren Siegel$^{c,}$\footnote{E-mail: siegel@insti.physics.sunysb.edu}}

\vskip .3cm
\centerline{$^a$\em Institute of Cosmology,
Department of Physics and Astronomy, Tufts University, Medford, MA 02155  }

\centerline{$^b$\em Department of Physics, University of California, Davis,
CA 95616}

\centerline{$^c$\em C.~N.~Yang Institute for Theoretical Physics, State
University of New York, Stony Brook, NY 11794}

\vskip 1.9cm

\begin{abstract}
We explore the relation between classical and quantum states in both open and 
closed (super)strings discussing the relevance of coherent states as a
semiclassical approximation. For the closed string sector a gauge-fixing of
the residual world-sheet rigid translation symmetry of the light-cone gauge is 
needed for the construction to be possible. The circular target-space loop 
example is worked out explicitly.
   
\end{abstract}

\end{titlepage}

\section{Introduction}

The motivation to investigate the semiclassical limit of fundamental strings 
is twofold. First, in view of the revived interest in the possibility of producing 
superstrings of cosmic size in models of brane inflation, it has been suggested that brane 
annihilation would leave behind a network of lower dimensional extended 
objects \cite{Tye, AlexGia, CMP, Polchinski} which would be seen 
as strings from the four dimensional point of view. This realization opens up the 
possibility of observing the cosmological consequences of cosmic strings, either 
from fundamental strings or from (wrapped) D-branes.  It is therefore interesting
to understand how to reconcile the usual quantum mechanical treatment of fundamental
strings with their expected classical behaviour at cosmological scales. Second, 
the semiclassical limit discussed in this paper may 
play a role in relation to the possible microscopic counting of 
states \cite{Sus, Sen, Str, Dab} associated with known classical supergravity solutions.

{}It is usually assumed that the description of a semiclassical string
state (a string of macroscopic size) is in terms of a coherent superposition 
of the fundamental string quanta. However, ealier attempts to build this type
of state in the covariant gauge quantization face serious difficulties
\cite{calucci}. In the next section we motivate the use of the lightcone gauge coherent state to 
give an accurate microscopic description of extended open string
solutions (spinning string configuration). We contrast the result with the
alternative  mass eigenstates (``perturbative states'') with the same angular momentum, 
emphasizing the advantages of the former. In section 3 we focus on the closed string case. 
We describe the obstacle to the naive extrapolation from the open string case.
We then provide the solution, suitable for the case of a circular target space
loop, by gauge-fixing the residual rigid $\sigma$ translation symmetry.
This is done in three ways: in unitary gauge, through a BRST method and in 
Gupta-Bleuler like quantization.

\section{Open strings}

In this section we explore the relation between perturbative quantum
states and their classical counterparts for open strings. In particular, we
will find the closest quantum state to the leading classical
Regge trajectory that corresponds to the following spinning configuration,
\beq\label{spinning}
X^0 &=& A\tau~, \nonumber\\
X^1 &=& A\sin \tau \cos \sigma~, \nonumber\\
X^2 &=& A\cos \tau \cos \sigma~.
\eeq
In the conventions of \cite{GSW}, the general solution for the open strings in
light-cone gauge is given by the following expressions,
\beq
X^+ &=&x^+ + l^2 p^+ \tau \nonumber~,\\
X^i &=& x^i + l^2 p^i \tau+
i l \sum_{n\not = 0} {1\over n}~\alpha_n^i~{\rm e}^{-i n \tau
}~\cos n\sigma~,\nonumber\\
X^-&=&x^-+l^2p^-\tau+il\sum_{n\not = 0}{1\over n}~\alpha_n^-~{\rm e}^{-in\tau
}~\cos n \sigma~.
\eeq
On the other hand, in order to fulfill the constraints the
$\alpha_n^-$ are restricted to be functions of the physical
transverse directions,
\beq
\alpha_n^- = {1\over {2 l p^+}} \left(\sum_{m=-\infty}^{\infty}
: \alpha_{n-m}^i \alpha_m^i : -2a\delta_n\right)~,
\eeq
where the $a$ coefficcient comes from the normal ordering of the
$\alpha$ operators. Furthermore we can also write the mass and the
angular momentum in terms of transverse modes in the following way,
\beq
M^2&=&{2\over l^2} \left(\sum_{n=1}^{\infty}\alpha_{-n}^i\alpha_n^i-a\right)~,
\\
J^{ij}&=&-i\sum_{n=1}^{\infty}{1 \over n}
\left(\alpha_{-n}^i\alpha_n^j -\alpha_{-n}^j\alpha_n^i\right)~.
\eeq

\subsection{Classical Regge trajectory in the light-cone gauge}

We start out by finding the description in the lightcone gauge of the 
classical solution of the open string found in (\ref{spinning}). It is 
clear that this state corresponds to a string spinning around its
center of mass, which has been set to the origin of the coordinates,
which implies that,
\beq
x^+=x^-=x^i=p^i=0~.
\eeq
Also, it is easy to see that the only excited oscillators
in this solution are,:
\beq
\alpha_1^1 &=& (\alpha_{-1}^1)^* = -{{i A}\over {2 l}}~,\\
\alpha_1^2 &=& (\alpha_{-1}^2)^* = {{ A}\over {2 l}}~.
\eeq
Which in turn implies that,
\beq
X^1 &=& il \left(-\alpha_{-1}^1 e^{i\tau} + \alpha_1^1
e^{-i \tau}\right) \cos \sigma = A \cos \sigma \cos \tau ~,\\
X^2 &=& il \left(-\alpha_{-1}^2 e^{i\tau} + \alpha_1^2
e^{-i \tau}\right) \cos \sigma = A \cos \sigma \sin \tau~.
\eeq
And for the $(-)$ oscillators we obtain,
\beq
\alpha_0^-= l p^- = {1 \over{2 l p^+}} \left({A \over l}\right)^2~,
\eeq
with all the other $\alpha_n^-=0$. Note that the only non-trivial
cases are $n = \pm 2$, which in our case are still zero,
\beq
\alpha_2^- = {1 \over{2 l p^+}} \left(\alpha_1^1 \alpha_1^1 + \alpha_1^2
\alpha_1^2 \right) =  {1 \over{2 l p^+}} \left[-\left({ A \over {2
      l}}\right)^2 + \left({ A \over {2 l}}\right)^2 \right] = 0~,
\eeq
and the same for $\alpha^-_{-2}$.
If we want the string to move on the $1-2$ plane only, we have to
impose that $X^9=0$,
\beq
X^9 = {1 \over {\sqrt{2}}} \left(X^+ - X^- \right) =  {1 \over {\sqrt{2}}}(l^2p^+ -l^2 p^-) \tau =0~,
\eeq
in other words, that $p^+=p^-=A/\sqrt{2} l^2$. This also means that,
\beq
X^0 =  {1 \over {\sqrt{2}}} \left(X^+ + X^- \right) = {1 \over {\sqrt{2}}} (l^2p^+ + l^2 p^-) \tau =
A \tau~.
\eeq
So, putting all these results together, we finally get a solution of the form (\ref{spinning}),
\beq
X^0 &=& A \tau ~,\nonumber\\
X^1 &=& A \sin \tau \cos \sigma~,\nonumber \\
X^2 &=& A \cos \tau \cos \sigma~.
\eeq
We can also use the equations given above to compute, in the classical
limit, the observables of this state, {\it i.e.}, its mass and angular
momentum.
\beq
M^2&=&{2\over l^2} \left(\sum_{n=1}^{\infty} \alpha_{-n}^i
\alpha_n^i\right)
=  {2\over l^2} \left( \alpha_{-1}^1 \alpha_1^1 +
\alpha_{-1}^2 \alpha_1^2 \right)  = {A^2 \over l^4}~,
\\
J^{12}&=& - i  \left( \alpha_{-1}^1 \alpha_1^2 -
\alpha_{-1}^2 \alpha_1^1 \right) = { A^2 \over {2l^2}}~.
\eeq
which indeed show that these configurations belong to the 
classical Regge trajectory of maximum angular momentum per
unit mass.

\subsection{Quantum State}

We will now try to obtain the quantum state for the open
bosonic string that resembles the classical case described
above. There seem to be two natural possibilities:

\subsubsection{``Perturbative'' Regge State (Mass Eigenstates)}

The classical calculation suggests that we construct the quantum state
for this solution in the following way,
\beq
|\psi\rangle = {1 \over \sqrt{2^n n !} } \left(\alpha_{-1}^2 - i
 \alpha_{-1}^1\right)^n |0\rangle~.
\eeq
The reason to choose this particular configuration becomes clear when we 
realize that this state is, in fact, an eigenstate of mass and angular momentum, 
\beq
M^2 |\psi\rangle &=&{2\over l^2}\left(\alpha_{-1}^1 \alpha_1^1 +
\alpha_{-1}^2\alpha_1^2-1\right)|\psi\rangle={2\over l^2} (n-1)|\psi\rangle~,\\
J^{12}|\psi\rangle &=&-i \left( \alpha_{-1}^1 \alpha_1^2 -
\alpha_{-1}^2 \alpha_1^1 \right)|\psi\rangle=n|\psi\rangle~.
\eeq
Using the identification $n = 1+A^2/ 2 l^2$, we see that
this state has identical values of mass and angular momentum to
the classical configuration in the $n \gg 1$ or $A \gg l$ limit.
In fact, they saturate the quantum inequality for mass eigenstates
given by,

\beq
J^{12} \le \frac{l^2}{2} M^2 +1
\eeq

On the other hand, this is not an eigenstate of the position of the string. 
It is easy to show that, in this state, the expectation value of the
spatial part of the string position operator is equal to zero for all values
of $\sigma$ and $\tau$, namely,
\beq
\langle\psi|\alpha_n^i|\psi\rangle&=&0~,\\
\langle\psi|\alpha_k^-|\psi\rangle&=& \langle\psi| {1\over {2 l p^+}}
\sum_{m=-\infty}^{\infty} \alpha_{k-m}^i \alpha_m^i |\psi\rangle = 0~,
\eeq
for $k \neq 0$. Finally,
\beq
\langle\psi|\alpha_0^-|\psi\rangle = {{n-1} \over{ l p^+}}~,
\eeq
which implies that\footnote{Where we have used the relations
  noted before between the different sets of parameters,  
$p^+ = {A \over {\sqrt{2} l^2}}$  as well as $n = {{A^2}\over {2 l^2}}+1 $.}
\beq
\langle\psi|X^0|\psi\rangle &=& A\tau~,\\
\langle\psi|X^i|\psi\rangle &=& 0~,\\
\langle\psi|X^9|\psi\rangle &=& 0~.
\eeq

We notice that while it is true that this state has similar properties 
to the classical one, it clearly does not resemble the macroscopic
string state, in the sense that its spacetime motion is not reproduced at all,
not even taking a high excitation number ({\it i.e.}, $n\gg 1$ does not
produce a semiclassical limit).

\subsubsection{``Coherent'' Regge State}

On the other hand, it seems more reasonable to try to mimic the classical configuration 
by constructing a coherent state of the form,
\beq
|\phi\rangle= e^{v\alpha_{-1}^1-v^*\alpha_{1}^1}~
e^{-iv\alpha_{-1}^2-iv^*\alpha_{1}^2}|0\rangle~,
\eeq
where $v$ is a parameter related to $p^+$, ($p^+=l^{-1}\sqrt{2|v|^2-1}$) and to the amplitude of
 the spacetime oscillations.

This state has the following expectation values for the energy and angular
momentum,
\beq
\langle\phi| M^2 |\phi\rangle&=&
{2\over l^2}\langle\phi|\left(\sum_{n=1}^{\infty}\alpha_{-n}^i
\alpha_n^i - 1 \right)|\phi\rangle \nonumber\\
&=& {2\over l^2} \langle\phi|\left( \alpha_{-1}^1 \alpha_1^1 +
\alpha_{-1}^2 \alpha_1^2 - 1 \right) |\phi\rangle={2\over l^2} (2|v|^2-1)~,\\
\langle\phi| J^{12}|\phi\rangle &=& - i  \langle\phi| \left( \alpha_{-1}^1 \alpha_1^2 -
\alpha_{-1}^2 \alpha_1^1 \right)  |\phi\rangle = 2|v|^2~,
\eeq
which also correspond to the values obtained in the previous section
by considering the identification $2|v|^2 = n$. The key point,
however, is that this state does have the spacetime position
expectation value of an extended string, namely,
\beq
 \langle\phi| X^0|\phi\rangle &=& l~\tau~\sqrt{4 |v|^2-2} = A~\tau~,\\
 \langle\phi| X^1|\phi\rangle &=&  2 l v~\sin \tau \cos \sigma =
 \sqrt{A^2+ 2 l^2}~\sin \tau \cos \sigma~,\\
 \langle\phi| X^2|\phi\rangle &=&  2 l v~\cos \tau \cos \sigma = \sqrt{A^2+ 2 l^2}~\cos \tau \cos \sigma~,\\
 \langle\phi| X^9|\phi\rangle &=&  0~,
\eeq
which for $A \gg l$ approaches the classical solution discussed
above.\footnote{Note that in this limit we also recover the same
 values for the mass and angular momentum of the classical
 configuration.} This shows that the
coherent state is a much closer match to the classical solution than
the previously considered construction. 

\section{Closed strings}

The calculations in the previous section show how one can obtain a
semiclassical coherent state for open strings in a very similar way to the simple 
harmonic oscillator. However, no such construction is available for 
closed strings, except in the approach of ``semiclassical''
quantization (in the sense of \cite{ls}, where the constraints
 are satisfied in mean value) as in \cite{bpi2}. In this section we will 
present the obstacle in finding a macroscopic state from perturbative closed 
string states in the lightcone gauge, and propose a solution.

\subsection{The problem of a microscopic perturbative description}

Consider the first order form of of the bosonic string action with world-sheet
coordinates $m=(0,1)\equiv(\sigma,\tau)$,
\beq\label{action}
S={1\over 2\pi \alpha^\prime}\int d^2\sigma \left(\partial_m X\cdot P^m+g_{mn}
{1\over 2}P^m\cdot P^n\right)~,
\eeq
where $g_{mn}=(-h)^{-1/2}h_{mn}$ is the unit determinant part of the world-sheet metric
$h_{mn}$ (related to the second order form \cite{act} using
$P^m=(-h)^{1/2}h^{mn}\partial_n X$).

Integrating out $P^\sigma$ via its equation of motion:
\beq
P^\sigma=-{1\over g_{11}}\left(X^\prime+g_{01}P^\tau\right)
\eeq
the action becomes
\beq\label{act2}
S={1\over 2\pi \alpha^\prime}\int d^2\sigma \left[\dot X \cdot P -
{1\over 2 g_{11}}\left(X^{\prime 2}+P^2\right)-
{g_{01}\over g_{11}}X^\prime\cdot P\right]~,
\eeq
where $P^\tau\equiv P$ to simplify notation.
The reparametrization symmetry of (\ref{action}) can be used to set $g_{11}=1$
and $g_{01}=0$. Further, Weyl symmetry can be used to set $h_{11}=1$ and
residual semilocal symmetry to set
\beq
X^+=\tau~,~~~~~P_+=1 ~~~~~~~~({\rm light-cone~ gauge})~.
\eeq
Let us now look closely at the constraint obtained by varying $g_{01}$,
{\it i.e.}, $X^\prime\cdot P=0$ that has the following mode decomposition
($\int_0^\pi~d\tau~{\rm e}^{2im\tau}\cdots$):
\beq
C_m&=&\alpha_m^--\t\alpha_m^-+\sum_{n\not = 0} \left(\alpha_{m-n}^i\alpha_n^i-
\t\alpha^i_{m-n}\t\alpha_n^i\right)~,
\eeq
where
\beq
C_0=\Delta N ~,~~~~~~~({\rm recall} ~~\alpha^-_0=\t\alpha_0^-)
\eeq
is the generator of rigid $\sigma$ shifts
($\delta X=\epsilon \partial_\sigma X$, with constant $\epsilon$). Upon
quantization,
if a physical state $|{\rm phys}\rangle$ satisfies the constraint $C_0$:
\beq
\Delta N |{\rm phys}\rangle=0~,
\eeq
then, it follows that
\beq
0\equiv\langle {\rm phys}|[\Delta N, X]
|{\rm phys}\rangle=\partial_\sigma\langle X \rangle~.
\eeq
This shows that the string would appear to be stuck at a fixed point $\langle X \rangle$ for all
$\sigma$, making it impossible, in this gauge, to have a
macroscopic extended closed string. The reason for this is that we have not
fixed the gauge completely so, in practice, we are integrating over all
the gauges compatible with the lightcone gauge which, of course, yields
the aforementioned center of mass position for the whole string. This makes the
operator $ X $ not the right quantity to look at in this
gauge if we are interested in evaluating the semiclassical position of the string. The
way out of this problem that we suggest in the following section is
to fix the gauge completely before evaluating the position of the string.

\subsection{Gauge-fixing $\sigma$ translations}
{}In this subsection we will fix the gauge for the residual rigid
reparametrization symmetry $\sigma\to
\sigma+\epsilon$ that remains after choosing light-cone gauge for closed
strings. We do this by prescribing the value of one of the coordinates modes 
of the string. 
This fixes the symmetry in a way similar to the way in which the open string 
arises from the closed string by removing the modes of one handedness 
\cite{s}.

{}Our goal here will be that of describing a circular string.
Once we are in the light-cone gauge, $X^+=(X^0+X^9)/\sqrt{2}\propto \tau$, we single out
two of the coordinates (that span the plane in which the circular loop lies):
$X^1$ and $X^2$. The solutions for the equations of motion of these
coordinates have the usual decomposition into 
left and right moving modes, namely, $X^i=X^i_L+X^i_R$.
\footnote{
For closed strings
the standard decomposition into left and right-movers is
\begin{eqnarray}
X^i&=&X^i_L+X^i_R~,\\
X^i_L&=&{1\over 2}x^i+{1\over 2} l^2 p^i(\tau-\sigma)+
{i\over 2}l
\sum_{n\not = 0} {1\over n} \alpha_n^i{\rm e} ^{-2in(\tau-\sigma)}~,\\
X^i_R&=&{1\over 2}x^i+{1\over 2} l^2 p^i(\tau+\sigma)+{i\over 2}l
\sum_{n\not = 0} {1\over n} \tilde\alpha_n^i{\rm e}^{-2in(\tau+\sigma)}~,
\end{eqnarray}
where $i=1, \cdots, 8$, and $x^i$ and $p^i$ are the center of mass
position and momentum of the loop.}

{}We propose the following additional gauge-fixing condition suitable for the 
description
of the circular string loop states:
\beq\label{GF}
\Phi={1\over\pi}\int d\sigma{\rm e}^{-2i\sigma}\left(\partial_-X^1
-v l {\rm e}^{-2i(\tau-\sigma)}\right)=0~,
\eeq
where $\partial_-=\partial_\tau-\partial_\sigma$ and the parameter $v$ will be 
related to the radius of the circle. Note that this is only a condition
involving the left-moving part of $X^1$ since $\partial_- X_R\equiv 0$.

To the action (\ref{act2}) we add the gauge-fixing term: 
\beq\label{Lgf}
{\cal L}_{gf}=\lambda \Phi~,
\eeq
such that $\lambda$ acts as the Lagrange multiplier enforcing the gauge conditions $\Phi=0$ 
({i.e.}, determining $\alpha_1^1$).

After reaching light-cone gauge, the only remaining piece of the
third term in the action (\ref{act2}) is given by %
\beq \int d\tau g_{0}\sum_{n}\left( \alpha_{-n}^i(\tau)
\alpha_n^i(\tau)-\t \alpha_{-n}^i(\tau)\t\alpha_n^i(\tau)\right)~,\eeq
where $g_0$ stands for the zero mode of $g_{01}$, we have used the
decomposition $\partial_-X^1=\sum \alpha_n^1(\tau) {\rm
e}^{2ni\sigma}$, $\dots$ etc. 

Varying with respect to $\lambda$ and $g_0$ we obtain the gauge-fixing 
condition and the constraint that can be solved (if $v\not =0$) for 
$\alpha_1^1(\tau)$ and 
$\alpha_{-1}^1(\tau)$ respectively. On the solutions, 
$\alpha_n(\tau)=\alpha_n {\rm e}^{-2in\tau},\dots$, etc. Therefore, we obtain,
\beq
\alpha_1^1&=&v~,\label{sol1}\\
\alpha_{-1}^1&=&-{1\over v}\left(\sum_{m\ge 2} \alpha_{-m}^1
\alpha_m^1+\sum_{n\ge 1} \alpha_{-n}^j
\alpha_n^j-\t \alpha_{-n}^i\t\alpha_n^i\right)~,\label{sol2}
\eeq
where $j=2,\cdots,8$. These results show that the idea behind our
gauge fixing choice in (\ref{GF}) is very much like the one used to
solve the constraints in the lightcone gauge expressing $\alpha_n^-$ 
in terms of the transverse modes. 

In this gauge, then, the mode decomposition of $X^1_L$ is different from the 
usual. It reads,
\beq\label{x1}
X^1_L&=&{1\over 2}x^i+{1\over 2} l^2 p^i(\tau-\sigma)
- {i\over 2}l~ \alpha_{-1}^1{\rm e} ^{2i(\tau-\sigma)}
+ {i \over 2} v~l~ {\rm e} ^{-2i(\tau-\sigma)}\nonumber\\ 
&&+ {i\over 2}l \sum_{n>1} {1\over n} 
\left(\alpha_n^1~{\rm e}^{- 2in(\tau-\sigma)}
-\alpha_{-n}^1~{\rm e} ^{2in(\tau-\sigma)}\right)~,
\eeq 
where $\alpha_{-1}^1$ should be interpreted as the operator on the rhs of (\ref{sol2}).

\vskip.4cm

Alternatively, using the BRST method, the same result can be
obtained with a gauge-fixing term linear in $\Phi$. The BRST charge
in this case contains an extra term:
\beq Q_{\rm extra}=c\lambda ~, \eeq
with corresponding ghost and anti-ghost $c$ and $b$ satisfying 
$\{c, b\}=1$.
  
The gauge-fixing term in this case is \beq {\cal L}_{gf}=\{Q,\Lambda\}~,
\eeq where
$\Lambda= b \Phi$ is the gauge-fixing function. The extra term $ Q_{\rm extra}$
gives the contribution (\ref{Lgf}), and there are also 
Fadeev-Popov terms from $b\{Q,\Phi\}$ with contribution 
from the term originally present in the BRST charge, $\tilde c \Delta N$. 
By using the 
gauge condition this contribution is $v\tilde c b$.
Thus, the ghosts decouple in this gauge.
 
Then, one proceeds as before, solving the gauge condition and constraint.

\vskip.4cm

Quantization in this gauge can also be achieved in a Gupta-Bleuler approach,
by imposing the gauge condition and constraint on physical states.  
For any pair of physical states
$|\chi\rangle$ and $|\phi\rangle$, we require: 
\beq\label{phi}
\langle\chi|\Phi|\phi\rangle=0~.
\eeq
But (\ref{phi}) is satisfied if we impose the following condition 
on physical states:
\beq\label{cond0}
\Phi|{\rm phys}\rangle=0~,
\eeq
Therefore, 
(\ref{cond0}) implies,
\beq\label{cond}
\alpha_1^1|{\rm phys}\rangle=v |{\rm phys}\rangle~,
\eeq

{}Using (\ref{cond}), the $\Delta N=0$ constraint can be
rewritten. Again, at the quantum level, for any pair of physical states
$|\chi\rangle$ and $|\phi\rangle$
\beq\label{DN}
0=\langle\chi|\Delta N|\phi\rangle~,
\eeq
should hold. But it is enough to impose:  
\beq
\left(v \alpha_{-1}^1+\sum_{m\ge 2} \alpha_{-m}^1
\alpha_m^1+\sum_{n\ge 1} \alpha_{-n}^j
\alpha_n^j-\t \alpha_{-n}^i\t\alpha_n^i\right)|{\rm phys}\rangle=0~,
\eeq
%


\subsection{The state}

Let us now consider a state in the gauge of the previous subsection of the 
following form
\beq
|\phi_0\rangle=|\phi_0\rangle_L\otimes|n\rangle_R~,
\eeq
where the left-moving factor is a coherent state built on a left vacuum,
$|0\rangle_L$,
\beq\label{state}
|\phi_0\rangle_L~=
{\rm e}^{-iv\alpha_{-1}^2-iv^*\alpha_1^2}|0\rangle_L~,
\eeq
and the right-moving part is an eigenstate of $N_R$ with eigenvalue $2
v^2$. We also set the parameters $x^i$ and $p^i$ to zero,
making the center of mass the origin of the coordinate system. 
Notice that $\alpha^2_1|\phi_0\rangle_L=-iv$ which implies that
$\langle\phi_0|\alpha^1_{-1}|\phi_0\rangle_L=v$ upon using equation (\ref{sol2}).

We can compute now the expectation value of the string coordinates in the
normalized state $|\phi_0\rangle$ to find a stationary circular loop of radius $vl$:
\beq
\langle X^0\rangle&=& 2l~\tau~\sqrt{2|v|^2-1},\nonumber\\
\langle X^1\rangle&=&{{il}\over 2}\langle\phi_0|
-\alpha_{-1}^1{\rm e}^{2i(\tau-\sigma)}+ 
v {\rm e}^{-2i(\tau-\sigma)}|\phi_0\rangle~,\nonumber\\
&=&v~l~ {\rm sin}~ 2(\tau-\sigma)~,\\
\langle X^2\rangle&=&{{il}\over 2}\langle\phi_0|
-\alpha_{-1}^2{\rm e}^{2i(\tau-\sigma)}+
\alpha_{1}^2{\rm e}^{-2i(\tau-\sigma)}|\phi_0\rangle~,\nonumber\\
&=&v~l ~{\rm cos}~ 2(\tau-\sigma)~.
\eeq
There is no contribution from the right-moving bosonic excitations to the 
expectation value because we are considering that this sector is in a $N_R$ eigenstate.
  
{}Before ending this subsection, let us note that the circular loop is just 
one possible coherent state constructed using this gauge.
It is not difficult to see the generalization to other shapes.
Take a left-moving component for the state of the form:
\beq
|\phi_0\rangle=
{\rm e}^{\cal A}|0\rangle_L~,~~~~~{\cal A}=\sum_{\{i,m\}\not =\{1,1\}}u_{mi}
\alpha_{-m}^i+u_{mi}^*\alpha_m^i~,
\eeq
$i=1,\cdots,8$ and $m$ runs over positive integers.
If the right-moving part of the state has level $N_R=n$, then
$\langle X^1\rangle$ is real provided
\beq
n-v^2=\sum_{\{i,m\}\not =\{1,1\}}|u_{mi}|^2~.
\eeq
Choosing the parameters $v$ and $n$ appropriately one can build a loop of 
arbitrary shape in target space. As a trivial example, consider $n=v^2$, 
$u_{mi}=0$ to obtain a folded string along the $X^1$ axis.

\section{Conclusions}

{}The use of coherent states allows for the construction of semiclassical 
states in superstring theory that bear close resemblance to the classical 
solutions. These are of relevance in both studies of some macroscopic defects
expected to arise in string theory descriptions of inflationary cosmology
and towards a microscopic entropy counting of certain string configurations 
that correspond to classical supergravity solutions.  

{}For the open string case we have shown that the coherent state
reproduces the classical motion in the target spacetime while other
'perturbative' excitations have large oscillations (averaging to zero)
around the center of mass position of the string. In the closed string 
case similar properties are found. Also, with the gauge-fixing of
section 3 we obtained the microscopic description of a left-moving
static loop supported by a right-moving world-sheet current; the
analog of a classical superconducting vorton solution \cite{vortons}. 

{}As a final remark let us mention that throughout the text we have used 
ten dimensional spacetime thinking about the bosonic part of a superstring, 
the modification to include the fermionic sector being straightforward. 
In particular, the state of subsection 3.3 could have fermionic right-moving 
excitations accounting for part or the whole of the level $N_R$ needed. 
The latter possibility leads to the loop completely stabilized by fermionic 
excitations \cite{bpi2} which has no classical gravitational radiation.

\section*{Acknowledgments}
We would like to thank Jaume Garriga, Ken Olum, Alex Vilenkin and
especially Roberto Emparan and Jorge Russo for illuminating 
discussions. AI is grateful to the organizers of the Simons Workshop at 
Stony Brook and Perimeter Institute for their hospitality while this 
work was in progress. The work of AI was supported by DOE Grant
DE-FG03-91ER40674. WS was supported in part by NSF Grant PHY-0354776.

\end{document}